\documentclass[intlimits,twoside,a4paper]{article}

\usepackage[cp1251]{inputenc}
\usepackage{pdfpages}
\usepackage[eqsecnum]{cmpj3}

\graphicspath{{Figures/}}

\issue{2023}{26}{3}{33301}
\doinumber{10.5488/CMP.26.33301}
\title[Vibrational spectroscopy in liquid water]%
{Vibrational spectroscopies in liquid water: on temperature and
  coordination effects in Raman and infrared spectroscopies%
}
\author[R. Vuilleumier, A.
P. Seitsonen]{R. Vuilleumier\orcid{0000-0001-5386-699X}\refaddr{ENS}, A.
  P. Seitsonen\orcid{0000-0003-4331-0650}\refaddr{UZH,ENS}\thanks{Corresponding
    author: \email{Ari.P.Seitsonen@iki.fi}.}}
\addresses{
\addr{ENS} Processus  d'Activation S\'{e}lective par
Transfert d'\'{E}nergie  Uni-\'{e}lectronique ou Radiatif (PASTEUR),
D\'epartement de Chimie,  \'{E}cole Normale Sup\'erieure, Paris Sciences et
Lettres (PSL) Research  University, Sorbonne Universit\'{e}, Centre
Nationale de la Recherche  Scientifique (CNRS), F-75005 Paris, France
\addr{UZH} Institut f\"{u}r Chemie, Universit\"{a}t Z\"{u}rich,
Winterthurerstrasse 190, CH-8057 Z\"{u}rich
}

\date{Received March 16, 2023, in final form April 25, 2023}
\begin{document}

\maketitle

\begin{abstract}
Water is an ubiquitous liquid that has several exotic and anomalous
properties. Despite its apparent simple chemical formula, its capability of
forming a dynamic network of hydrogen bonds leads to a rich variety of
physics. Here we study the vibrations of water using molecular dynamics
simulations, mainly concentrating on the Raman and infrared spectroscopic
signatures. We investigate the consequences of the temperature on the
vibrational frequencies, and we enter the details of the hydrogen bonding
coordination by using restrained simulations in order to gain quantitative
insight on the dependence of the frequencies on the neighbouring
molecules. Further we consider the differences due to the different methods of
solving the electronic structure to evaluate the forces on the ions, and report
results on the angular correlations, isotopic mixtures HOD in H$_2$O/D$_2$O and
and the dielectric constants in water.

%
\printkeywords
%
\end{abstract}

  \newcommand{\labelZZZ}{A$_0$D$_{00}$} \newcommand{\labelZZO}{A$_0$D$_{01}$}
  \newcommand{\labelZOO}{A$_0$D$_{11}$} \newcommand{\labelOZZ}{A$_1$D$_{00}$}
  \newcommand{\labelOZO}{A$_1$D$_{01}$} \newcommand{\labelOOO}{A$_1$D$_{11}$}
  \newcommand{\labelTZZ}{A$_2$D$_{00}$} \newcommand{\labelTZO}{A$_2$D$_{01}$}
  \newcommand{\labelTOO}{A$_2$D$_{11}$} \newcommand{\labelTOOs}{A$_2$D$_{11}$s}
  \newcommand{\labelHnm}{A$_3$D$_{nm}$}  \newcommand{\labelHOO}{A$_3$D$_{11}$}


%
%
\section{Introduction}\label{introduction}

 The vibrational properties of liquid water are already well known --- there is of course a plethora of literature on the topic, for a recent review please see for 
 example~\cite{Perakis_2016_ChemRev_a}.
The methods mostly used to characterise
them are the infrared (IR) and Raman spectroscopies in the local frame, and
inelastic neutron and X-ray scattering for the collective modes. The spectra
consist of high-frequency region, that stems from the intramolecular O-H
stretching motions, intramolecular H-O-H bending mode and the low-frequency
modes plus possibly combination modes and overtones, that arise from the motion of the water molecules with respect to each
other. The details of the vibrations change according to the instantaneous
environment of the molecules, in particular, the hydrogen bond (HB) network,
that forms as a ``glue'' between the molecules in the liquid. It is the
influence of these HBs on the vibrational frequencies that we want to
characterise in the present work.

An interesting temperature dependency of the signal in the O-H stretching bond
region in (IR) and Raman vibrational spectroscopies is known: the intensity
shifts to higher frequencies upon heating of the system. This can be understood
in terms of the changing coordination in the HB network formed between the
molecules: at low temperatures more HBs are present and thus the vibrational
frequencies are lowered due to the attractive interactions along the O-H
intramolecular bonds, causing the bond length to elongate and thus lower the
frequency. At higher temperatures, many of the HBs are lost and the molecules
approach the higher vibrational frequencies of a free water monomer. Thereby
the vibrational frequencies at varying temperatures are indirectly functions of
the intermolecular HB network, and thus the HB coordination of a given
molecule.

The computation of IR and Raman spectra directly from the electronic structure
in the condensed phase was developed along the appearance of ``modern theory of
polarization''~\cite{Resta_2010_a}. It was soon applied to simulation of the IR
spectra of water by Parrinello and co-workers~\cite{Silvestrelli_1997_CPL_a},
and soon the implementation of Raman spectra followed together with an
application in ice~\cite{Putrino_2002_a}.

Here we present atomistic molecular dynamics (MD) simulations of the
temperature dependency of the Raman and IR signals. The first studies of Raman
spectra directly from the electronic structure appeared some time
ago~\cite{Wan_2013_a}. Furthermore, we evaluate the Raman signal from the MD
trajectories recently simulated by Del Ben and co-workers using the MP2
\textit{ab initio} electronic structure methods to evaluate the interactions
between the water molecules~\cite{DelBen_2015_a}. In order to fully
characterize the underlying HB network, we first provide an analysis of the
structural and dynamical properties of the samples. We then proceed to
analysing the simulated spectra, and divide the changes detected to their
origins in the HB network.

We further note that nowadays it would be more fashionable to run the
simulations performed here using the potentials and required quantities such as
the polarizability developed via machine learning~\cite{Sommers_2020_PCCP_a}
rather than running them with the explicitly resolved electronic structure at
each time step. This approach could be used to reduce the errors due to the
needy sampling of the statistics, nuclear quantum effects, and one is able to apply
more accurate methods in solving the electronic structure and so on.

The present work is organized as follows: in  section~\ref{section:methods}
we present the methods of simulation and analysis and the notation, in 
section~\ref{section:results} we present the results obtained from our analysis
and in  section~\ref{section:discussion} we put our results into perspective
of what is already known about the vibrational spectroscopy of water and what
we have learnt here.

We already humbly apologize here for the omissions that we cannot cover all the
vast, sometimes even contradictory, literature and models and results present
in the literature from about a century of scientific discoveries.

%
%
\section{Methods}\label{section:methods}

We have performed density functional theory-based molecular dynamics (DFTb-MD),
where the ions are moved according to the Newtonian equations of motion
according to the forces derived from the density functional theory
(DFT)~\cite{Hohenberg_1964_PR_a}. The nuclei, in particular, the protons --- with
the neutron when simulating with deuterium --- in the hydrogen atoms, are
treated as point particles and we rely on the Born-Oppenheimer
approximation. Thus, the dynamics is performed using classical dynamics, leading
to non-quantized dynamics and, therefore, we should expect to obtain exactly the
same results as in the results from the experiments, added to the other
approximations that we are employing.

%
%
\subsection{Systems}\label{section:systems}

Our samples consisted of 128 water molecules in a periodically repeated cell;
in the MP2 simulations, 64 water molecules had been used
instead~\cite{DelBen_2015_a}. The different systems simulated are listed in
table~\ref{table:systems}. For computational simplicity we used the same,
ambient experimental density in all the simulations. In the DFTb-MD simulations
results on the angular correlations, isotopic lattice constant  of isotropic mixtures was
15.7459~{\AA}, corresponding to a density of 0.9808~g/cm$^3$ of light or
1.0904~g/cm$^3$ of deuterated water.

\begin{table*}[!t]
\caption{Summary of the system parameters: targeted temperature ($T$), the
  resulting average intramolecular O-H bond length $d_\mathrm{O-H}$, average
  band gap, dipole moment $\mu_\text{W}$ and average polarizability
  $\overline{\alpha}$, diffusion constant $D_\text{O}$ of O evaluated via
  the Einstein (E) or Green-Kubo (GK) formalism. XC(traj) is the approach used
  in generating the trajectory and XC(pol) in calculating the band gap, dipole
  moment or polarizability.}
\vspace{0.9mm}
\label{table:systems}
\begin{center}
\begin{tabular}{ll|ccccccccc}\hline\hline
  & $T$ & $d_\mathrm{O-H}$ & gap & $\mu_\text{W}$ & $\overline{\alpha}$ & $D_\mathrm{O}^\mathrm{E}$ & $D_\mathrm{O}^\mathrm{GK}$ \strut\\
 Notation & [K] & [\AA] & [eV] & [D] & [\AA$^3$] &
  \multicolumn{2}{c}{[\AA$^2$/ps]} \strut\\\hline
  \multicolumn{8}{c}{\bf XC: BLYP+D3 128 H$_2$O/D$_2$O}\strut\\
D$_2$O--$T=25$\textcelsius{}        & 298.15 & 0.992 & 4.39 & 3.03 & 1.771 & 0.08 & 0.10\strut\\ 
D$_2$O--$T=50$\textcelsius{}        & 323.15 & 0.990 & 4.29 & 2.96 & 1.767 & 0.21 & 0.18\strut\\ 
D$_2$O--$T=75$\textcelsius{}        & 348.15 & 0.990 & 4.18 & 2.93 & 1.765 & 0.29 & 0.23\strut\\ 
D$_2$O--$T=100$\textcelsius{}       & 373.15 & 0.989 & 4.12 & 2.89 & 1.761 & 0.40 & 0.37\strut\\ 
D$_2$O--$T=125$\textcelsius{}       & 398.15 & 0.988 & 4.02 & 2.85 & 1.760 & 0.51 & 0.59\strut\\ 
H$_2$O--$T=50$\textcelsius{}        & 323.15 & 0.992 & 4.30 & 2.99 & 1.769 & 0.17 & 0.21\strut\\ 
H$_2$O@D$_2$O--$T=50$\textcelsius{} & 323.15 & 0.992/0.991       & 4.31 & 3.03 & 1.769 & 0.07/0.08 & 0.14/0.13\strut\\ 
D$_2$O@H$_2$O--$T=50$\textcelsius{} & 323.15 & 0.991/0.991       & 4.31 & 3.05 & 1.767 & 0.39/0.17 & 0.07/0.13\strut\\ 
HOD@D$_2$O--$T=50$\textcelsius{}    & 323.15 & 0.991/0.991/0.992 & 4.32 & 2.98 & 1.774 & 0.28/0.17 & 0.11/0.15\strut\\ 
HOD@H$_2$O--$T=50$\textcelsius{}    & 323.15 & 0.992/0.991/0.990 & 4.28 & 2.99 & 1.769 & 0.16/0.22 & 0.08/0.14\strut\\ 
\hline
  \multicolumn{8}{c}{\bf XC: revPBE+D3 128 D$_2$O}\strut\\
D$_2$O--$T=25$\textcelsius{}        & 298.15 & 0.986 & 4.60 & 2.94 & 1.726 & 0.15 & 0.13\strut\\ 
D$_2$O--$T=50$\textcelsius{}        & 323.15 & 0.986 & 4.41 & 2.88 & 1.722 & 0.26 & 0.22\strut\\ 
D$_2$O--$T=75$\textcelsius{}        & 348.15 & 0.985 & 4.34 & 2.85 & 1.716 & 0.35 & 0.41\strut\\ 
\hline
  \multicolumn{8}{c}{\bf XC(traj): vdW-DF-rPW86/XC(pol): BLYP 128 H$_2$O}\strut\\
H$_2$O--vdW-DF/BLYP         & 323.15 & 0.984 & 4.00 & 2.80 & 1.751 & 0.34 & 0.37\strut\\ 
\hline
  \multicolumn{8}{c}{\bf XC(traj): MP2/XC(pol): BLYP 64 H$_2$O}\strut\\
MP2/BLYP                    & 298.15 & 0.978 & 4.67 & 3.02 & 1.667 & 0.07 & 0.08\strut\\ 
  \multicolumn{8}{c}{\bf XC(traj): MP2/XC(pol): PBE  64 H$_2$O}\strut\\
MP2/PBE                     & 298.15 & 0.978 & 4.67 &      & 1.685 & & \strut\\ 
\hline\hline
\end{tabular}
\end{center}
\end{table*}

We ran most of the simulations on the deuterated water, in order to reduce the
error due to the employment of the classical equations of motion without
quantum nuclear effects on the nuclei, and at somewhat elevated temperature of
50\textcelsius{}, since it is known that some of the methods used lead to samples that
are near the amorphous liquids close to the real melting point of the liquid, in
particular the BLYP+D3 that is described below.

%
%
\subsection{Computational details}\label{section:computational_details}

We have carried out DFTb-MD calculations using a hybrid Gaussian plane-wave
(GPW) method~\cite{lippert_hybrid_1997} as implemented in the QuickStep
module~\cite{VandeVondele_2005_a} in the \texttt{CP2K} suite of
programs~\cite{CP2K}. This approach combines a Gaussian basis set for the wave
functions with an auxiliary plane wave basis set for the density. We chose a
triple-zeta valence doubly polarised (TZV2P) basis set since it has been shown
to provide a good compromise between accuracy and computational
cost~\cite{VandeVondele_2005_b}.  The cut-off for the electronic density was
set to 400 Ry, in conjunction with the use of a smoothing method (NN50) for the
exchange-correlation contribution~\cite{VandeVondele_2005_a}. The smoothing
method has been found essential in our recent study of
water~\cite{Jonchiere_2011_a} to obtain converged forces.

The action of core electrons on the (pseudo) valence orbitals was replaced by
the Goedecker-Teter-Hutter (GTH) norm-conserving
pseudo-potentials~\cite{Goedecker_1996_a,Hartwigsen_1998_a}. The BLYP
exchange-correlation (XC) functional~\cite{Becke_1988_a,Lee_1988_a} of type
generalised gradient approximation (GGA) has been used together with
semi-empirical additional three-body potential~\cite{Grimme_2010_a} to account
for the dispersion interactions missing in the
GGA~\cite{Lin_2009_JPCB_a}. Improvement in the simulation of neat water was
already  shown in our previous study~\cite{Jonchiere_2011_a}. Since at the
time of the evaluations of the polarizabilities, dipole moments and band gaps,
the usage of MP2 and vdW-DF was not available or would have been prohibitively
expensive computational task, we used either BLYP or the alternative GGA,
Perdew-Burke-Ernzerhof (PBE)~\cite{Perdew_1996_a}, to evaluate the electronic
structure; the approximation of XC used is indicated as ``XC(traj)'' in
table~ref{table:systems}. Further, we simulated water using an explicit van der
Waals-density functional; we applied the original vdW-DF~\cite{Dion_2004_a}
otherwise but the revised Perdew-Wang-86 (rPW86)
approximation~\cite{Murray_2009_a} in the exchange part, as was done also in
the vdW-DF2 functional~\cite{Lee_2010_a}.

Born-Oppenheimer molecular dynamics simulations were carried out in the
canonical, or the NVT ensemble with a Nos\'{e}-Hoover thermostat chain and a
time step of 0.5 fs.  Trajectories were run for at least 50~ps, the light water
simulation H$_2$O--BLYP+D3 for over 150~ps. The properties were analyzed after
an equilibration section of 10 ps. We remind the Reader here that the nuclear
quantum effects have not been treated in the present study.

We characterize the HB network around a molecule by its HB coordination
$C_\mathrm{HB}$, and we use the notation A$_l$D$_{nm}$ when the central
molecule accepts $l$ HBs and donates $n$ and $m$ HBs from the two hydrogen
atoms on it. By symmetry, A$_l$D$_{nm}$ corresponds to A$_l$D$_{mn}$.

In order to observe a stable coordination of the HBs on a single molecule, we
constrained the local hydrogen bonding using restraints on collective variables
on the coordination number of the form
\begin{displaymath}
 C_{\mathrm{L}_1,\mathrm{L}_2} = \frac 1{\mathrm{N}_{\mathrm{L}_1}}
\sum_{j=1}^{\mathrm{N}_{\mathrm{L}_1}} \Bigl\{
\sum_{i=1}^{\mathrm{N}_{\mathrm{L}_2}} \frac
    {1-\left(r_{ij}/r_0\right)^n}{1-\left(r_{ij}/r_0\right)^m} \Bigr\},
\end{displaymath}
where $\mathrm{L}_2$ is one of the two hydrogens or the oxygen atom on the
central, constrained molecule and $\mathrm{L}_1$ is either any oxygen or
hydrogen in the sample, correspondingly. We used a harmonic potential between
the target value (number of hydrogens around a central oxygen or vice versa)
and $C_{\mathrm{L}_1,\mathrm{L}_2}$. The trajectory ``A$_2$D$_{11}$s'' is like
the ``A$_2$D$_{11}$'' but the radius $r_0$ is smaller, forcing stronger the
HBs; thus, ``s'' for ``strict''. In ``A$_3$D$_{nm}$s'' only the oxygen is
constrained whereas the hydrogens are not restrained to form or to avoid HBs.
We are aware that we only constrain the distance between the hydrogen and
oxygen atoms, and there is no enforcement of a suitable directionality in the
O-H$\cdots$O angle. Moreover, due to the relatively large radii used, this scheme
does not necessarily lead to a definite hydrogen bond or the hindering of one
because we do not wish to affect the vibrational modes strictly by forcing
caging. Rather the absolute values and shifts should be regarded as trends.

%
%
\subsection{Methods of analysis}\label{section:methods_of_analysis}

There was used a geometrical definition of an HB, that was formed between two molecules when the
O$\cdots$O distance is at most 3.5~\AA{} and the H$-$O$\cdots$O angle is
smaller than 30$^\circ$. 

The vibrational modes were analyzed by localizing them on the central,
restrained molecule using the effective normal mode analysis (ENMA)
method~\cite{Martinez_2006_a}.

The evaluation of the polarizability~\cite{Luber_2014_a} and the dipole moment
was done using the correponding implementations in the \texttt{CP2K} suite.

The dielectric constants were calculated using the formulae
\begin{eqnarray}
  &&\frac {\varepsilon_\infty-1}{\varepsilon_\infty+2} = \frac
        {4\piup\alpha}{3\Omega},
\nonumber\strut\\
  &&\varepsilon = \varepsilon_\infty + \frac
              {\left\langle{}M^2\right\rangle-\left\langle{}M\right\rangle^2}
              {3\varepsilon_0 \, k_\mathrm{B}T \, \Omega} \ , \nonumber
\end{eqnarray}
where $\alpha$ is the polarizability, $\Omega$ is the volume of the super cell and
$M$ is the total dipole moment. We estimate the latter from
the Berry phase of the dipole operator.

The frequencies from the MP2 trajectories were scaled with a uniform factor
0.94, similar to \cite{DelBen_2015_a}, to allow an easier comparison with
the other data. All the power, IR and Raman spectra were multiplied with the
frequency in resemblance to incorporating, and as an \textit{ad hoc} correction
due to, the quantum nature of the nuclei. 

%
%
\section{Results and discussions}\label{section:results}

%
%
\subsection{Structure and dynamics}\label{section:structure_dynamics}

\begin{figure}[!t]
\begin{center}
\includegraphics[width=0.6\columnwidth]{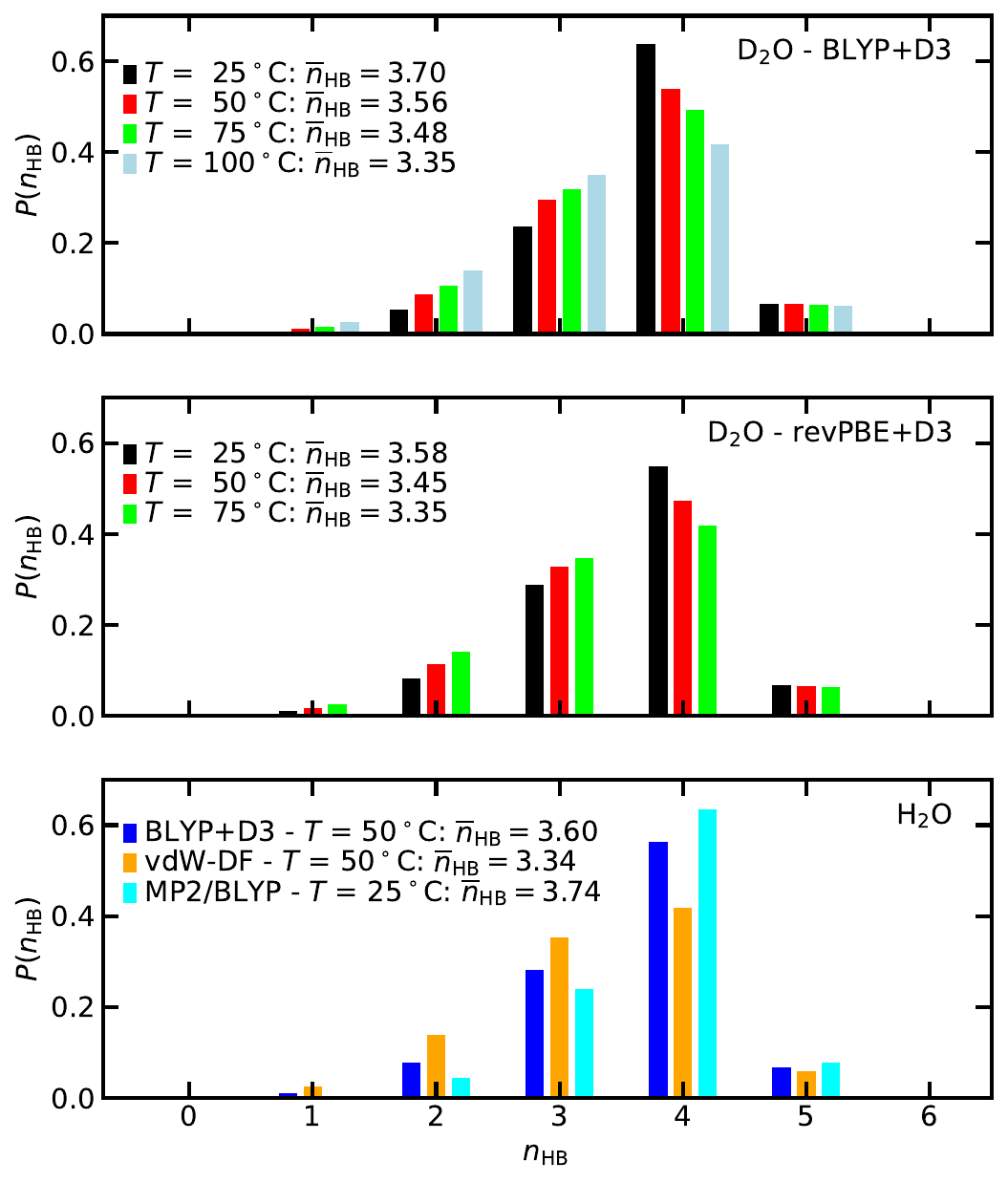}
\end{center}
\caption{(Colour online) Distribution of $P(n_\mathrm{HB})$ and average
  $\bar{n}_\mathrm{HB}$ number of HBs per molecule.}
\label{figure:nHB}
\end{figure}

We characterized the geometrical structure of the liquid samples via the
average number of HBs and intramolecular O-H bond length $d_\mathrm{O-H}$,
radial distribution functions (RDF) $g(r)$, the dynamics with the diffusion
constants, and the electronic properties by the average electronic band
gap,\footnote{We remark that the band gap is evaluated from differences in {K}ohn-{S}ham
	eigenvalues that i) are not guaranteed to bear any physical significance in
	the case of conduction band minimum and ii) are evaluated using a {GGA}
	approximation to the {XC}, which is known to yield a too small energy
	difference; yet here we are only interested in the changes between different
	systems, which are very likely to be reliable.}
 molecular dipole moment and polarizability; the
results are shown in table~\ref{table:systems} and figures~\ref{figure:nHB} and
S1 [here and below, prefix ``S'' refers to the Supporting Material (SM)]. 

Counter-intuitively, the average intramolecular bond length becomes slightly
shorter upon increasing the temperature, but this is probably due to the
concurrent reduction in the HBs. There are fewer HBs in the trajectory
simulated with the vdW-DF, and the $d_\mathrm{O-H}$ is shorter than with
BLYP+D3. MP2 yields the shortest O-H bond lengths, but the number of HBs is
similar to the results with BLYP+D3 also at ambient temperature.

Except for $d_\mathrm{O-H}$ all the other studied quantities follow the
expected trend upon an increase of the temperature: the number of HBs is reduced, even if the number of five-fold HB-coordinated molecules remains practically constant, the diffusion is enhanced, the electronic band gap is decreased due to larger geometrical fluctuations, the dipole moment decreases from the larger value found in ice and the polarizability is diminished. 


In the RDF, the closest reminescense between the BLYP+D3 and the MP2 simulations
is with the former simulated at $T=25$\textcelsius{}: the height of the peaks is very
similar both in O-O and O-H correlations, but the position of the first maximum
$r_\mathrm{O-O}^\mathrm{max;1}$ in $g_\mathrm{O-O}$ is at somewhat smaller
radius in the MP2 results. This comparison is, however, somewhat blurred by the
densities used: in the MP2 simulation the equilibrium density --- about 2\% larger than experimentally --- had been used, whereas
the BLYP+D3 samples are under pressure, the
equilibrium density at $T=295$~K, $P=1$~bar being
1.066~g/cm$^3$~\cite{DelBen_2015_a}. The simulation applying the vdW-DF yields
much less structured liquid than MP2 at $T=25$\textcelsius{} or even BLYP+D3 at
$T=100$\textcelsius{}, and $r_\mathrm{O-O}^\mathrm{max;1}$ is larger than even with
BLYP+D3. Interestingly, data of the fifth-nearest-neighbour RDF, shown in figure~S3, indicate that the fifth-closest O atom
to the central O~--- as important indication of the liquid-like behaviour in
water --- does not get closer at the minimum encounters, but the relative
magnitude at the maximum increases, as expected when the liquid becomes more
diffusive upon higher temperature. 

The electronic band gap gives a hint of the intra- and intermolecular
arrangement, now that it was evaluated with the same treatment of the XC in all
the computations of the Kohn-Sham eigenvalues. Therefore, it is interesting to
notice that the value from the vdW-DF trajectory is clearly smaller, beyond the
fluctuations (root-mean-square deviation $\approx$0.07~eV), than in the BLYP+D3
simulations at the same temperature, $T=50$\textcelsius{}.

We extract the values 1.964, 1.961, 1.960, 1.957, 1.956 for the dynamic
dielectric constant with the BLYP+D3 approach in the D$_2$O at the different,
increasing temperatures, 1.963 H$_2$O--$T=50$\textcelsius{} with the BLYP+D3, 1.932,
1.929, 1.925 at the temperatures 25, 50 and 75\textcelsius{} with the revPBE+D3,
1.950 with H$_2$O--vdW-DF/BLYP, and 1.949 and 1.963 in MP2/BLYP and MP2/PBE,
respectively. On the dielectric constant $\varepsilon$ we extract a value of
about 58, clearly but consistenly smaller than the value of $112\pm{}6$
obtained with the PBE approach and 64 water molecules at
350~K~\cite{Zhang_2016_JPCL_a}. Details on the derivation of our value is given
in the SM.

The diffusion constants $D_\mathrm{O}^\mathrm{E}$ and
$D_\mathrm{O}^\mathrm{GK}$ have somewhat different values, but the differences
are well within the statistical fluctuations, since much longer simulations should
be performed to achieve a better convergence. The main effects, an increase in
diffusion with the temperature in D$_2$O and respectively slower and faster
diffusion with MP2 and vdW-DF than with BLYP+D3 in H$_2$O, are expected based
on the relative magnitudes around the first minimum in the RDFs.

%
%
\subsection{Dependence on temperature and XC}\label{section:purewater-temperature*}

First we discuss the dependency of the Raman and infrared spectrum on the
temperature, and various treatments of the electronic structure problem.

Before presenting the simulated spctra, we briefly discuss the power spectrum,
that underlie the IR and Raman spectra. They are shown in figures~S13 and S14,
evaluated from the velocities of hydrogen or deuterium and the oxygen atoms in
H$_2$O and D$_2$O. As the temperature is increased in D$_2$O, there are several
systematic shifts along temperature: toward lower frequencies in regions around
300 and 650~cm$^{-1}$ and toward higher frequencies in the O-H stretching
region around 2300 and 2600~cm$^{-1}$, where the latter effect is the most
visible one and is seen clearly even in the power spectrum of O. The bending
frequency is highly unaffected by the varying temperature, only via the height,
which is set by normalizing the maximum of the whole spectrum and occurs in the
O-H stretching region. 

In H$_2$O, the vdW-DF- and MP2-treatments of the electronic structure lead to
much narrower spectrum than BLYP+D3 in the O-H stretching region; we remind
that the MP2 spectrum is multiplied by a constant whose value brings the
maximum of the O-H stretching to the same region as BLYP+D3. Thereby, the
maximum of the H-O-H stretching frequency is somewhat lower with MP2 than with
BLYP+D3, but the spectra are very similar in region 300-1000~cm$^{-1}$. At
around 280~cm$^{-1}$, vdW-DF has much lower intensity than BLYP+D3 or MP2, and
at around 120~cm$^{-1}$ MP2 simulations yield a lower magnitude than BLYP+D3 or
vdW-DF.

\subsubsection{Raman spectra}\label{section:purewater-Raman}

\begin{figure*}[!b]
\begin{center}
\begin{tabular}{c}
\includegraphics[width=0.75\textwidth]{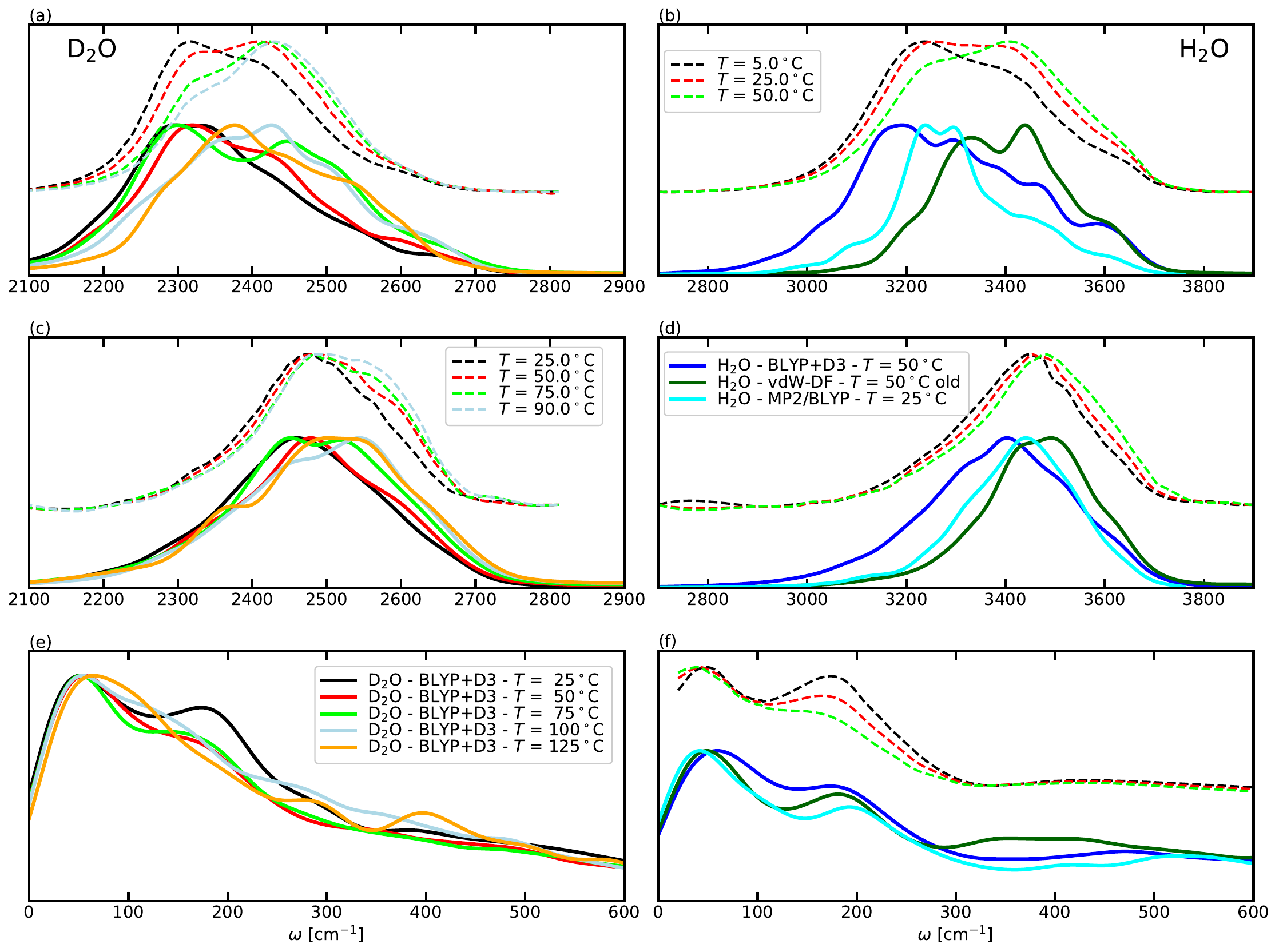}
\end{tabular}
\end{center}
\caption{(Colour online) a), b) Isotropic and c) -- f) anisotropic Raman spectra in
  a), c), e) D$_2$O and b), d), f) H$_2$O at different temperatures
  from the DFTbMD simulations together with experimental data, that
  were interpolated at the given temperatures [D$_2$O, original
    experimental data from Scherer {et
      alia}~\cite{Scherer_1974_a} at different temperatures from $-10$
    to $+90$\textcelsius{} in steps of 20 degrees]; H$_2$O, from Paolantoni
  {et al.}~\cite{Paolantoni_2007_a} --- low frequencies at
  three different temperatures --- and Scherer {et al.}~\cite{Scherer_1974_a} at different temperatures from $-10$
  to $+90$\textcelsius{} in steps of 20 degrees].}
\label{figure:spectra_Raman}
\end{figure*}

The evaluated isotropic and anisotropic Raman spectra in D$_2$O and
H$_2$O are presented in figures~\ref{figure:spectra_Raman} together
with experimental results~\cite{Scherer_1974_a,Paolantoni_2007_a}. In
the O-H stretching region, there is a clear decrease in intensity at the
lower frequency region and a corresponding increase in the higher
frequency region, both in isotropic and anistropic spectra. In the
low-frequency range, there are two peaks at 45 and 175~cm$^{-1}$,
where the relative intensity of the latter decreases with the
temperature.

The simulated spectra  follow these trends reasonably well: overall the width of
the O-H stretching band from BLYP+D3 is larger than in experiments, more so in
the isotropic spectrum. In D$_2$O we have scaled \textit{ad hoc} the
experimental frequencies by 0.97 to make a comparison easier. In
figure~\ref{figure:spectra_Raman}, we have colour-coded the experimental results
somewhat differently than in BLYP+D3 results, by visually adjusting the
experimental temperature that corresponds to the BLYP+D3 temperature most
adequately. We see that the BLYP+D3 intensities roughly correspond to
experimental data at 25 degrees lower temperature, for example
BLYP+D3-$T=25$\textcelsius{}  correspond to experiment-$T=0$\textcelsius{}. These deviations from
experimental results mostly originate from the inaccuracy of the BLYP+D3
treament of the electronic structure, possibly having a contribution due to the
nuclear quantum effect neglected here. In BLYP+D3-$T=75$\textcelsius{} spectra, there
is a double-peak structure close to the maximum that does have a corresponding
feature in the experimental counter-part. This might originate from the
relatively short simulation time into the trajectory, 40~ps. In the
low-frequency part, the relative intensity of the peak at about 175~cm$^{-1}$
does not decrease exactly monotonously in BLYP+D3; at 100\textcelsius{}, the peak is
no longer present in the BLYP+D3 spectrum.

In H$_2$O, the comparison of isotropic spectra from BLYP+D3 with experimental ones
is similar, with the simulated spectra at $T=50$\textcelsius{} being now similar to the
experimental one obtained at $T=5$\textcelsius{}; here, the experimental frequency
scale is not scaled. In the spectrum from vdW-DF/BLYP, the intensity is at
higher frequencies, corresponding to the difference seen in the relative power
spectra; the simulated spectra at $T=50$\textcelsius{} now resemble the experimental
one obtained at $T=50$\textcelsius{}, being in better agreement in the absolute
frequency, but the width of the whole O-H band is narrower and the distance
between the two peaks is smaller than in the experimental spectra. The low-frequency
peaks in the anisotropic spectra are well reproduced both in BLYP+D3 and
vdW-DF/BLYP-derived results. The anisotropic spectra in the O-H stretching region
are similar to the experimental ones, with a difference in the
frequency scale similar to the isotropic case.

The simulated Raman spectra are very similar to each other when either BLYP- or
PBE-GGA was used to calculate the polarizability in the MP2 trajectory, and we
only show the former here. The isotropic spectrum is much narrower than in the
experiments, but over-all the agreement is best with the shape of the
experimental spectrum measured at $T=5$\textcelsius{}. The relative intensity of the
low-frequency peaks is similar to the experimental ratio at $T=25$\textcelsius{}; the
shape of the high-frequency spectrum agrees well with the experimental one.

\subsubsection{infrared spectra}\label{section:purewater-IR}

\begin{figure*}[!t]
\begin{center}
\begin{tabular}{c}
\includegraphics[width=0.7\textwidth]{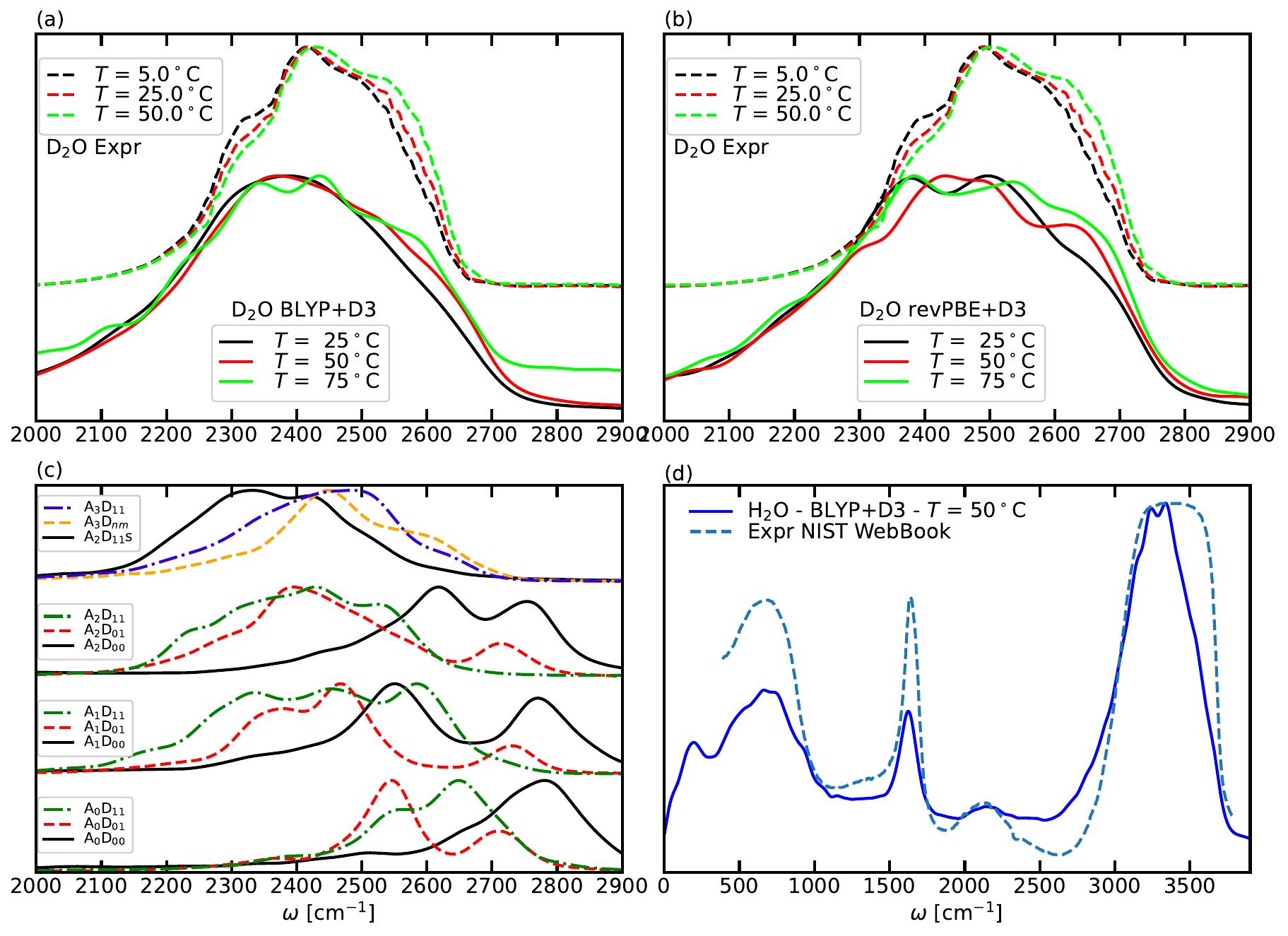}
\end{tabular}
\end{center}
\caption{(Colour online) infrared signal calculated in a) in D$_2$O from BLYP+D3 and
  experiments -- dashed lines, b) in D$_2$O from revPBE+D3 and experiments, c)
  the constrained HB coordinations in D$_2$O and d) in H$_2$O. Experimental
  data from~\cite{Marechal_2011_JMolStruc_a} in Panels (a) and (b) and from~\cite{NIST_WebBook} in panel (d).}
\label{figure:IR_spectra}
\end{figure*}

The simulated IR spectra are shown in figure~\ref{figure:IR_spectra}. As in the
case of Raman spectroscopy, experimental results in the IR spectra in the O-H
stretching region pose a shift of intensity from lower to higher frequencies
upon increasing temperature. This change in intensity is well reproduced in the
BLYP+D3 simulations. In H$_2$O, there is a double-peak structure.

\begin{figure*}[!b]
  \begin{center}
    \begin{tabular}{c}
      \includegraphics[width=0.6\textwidth]{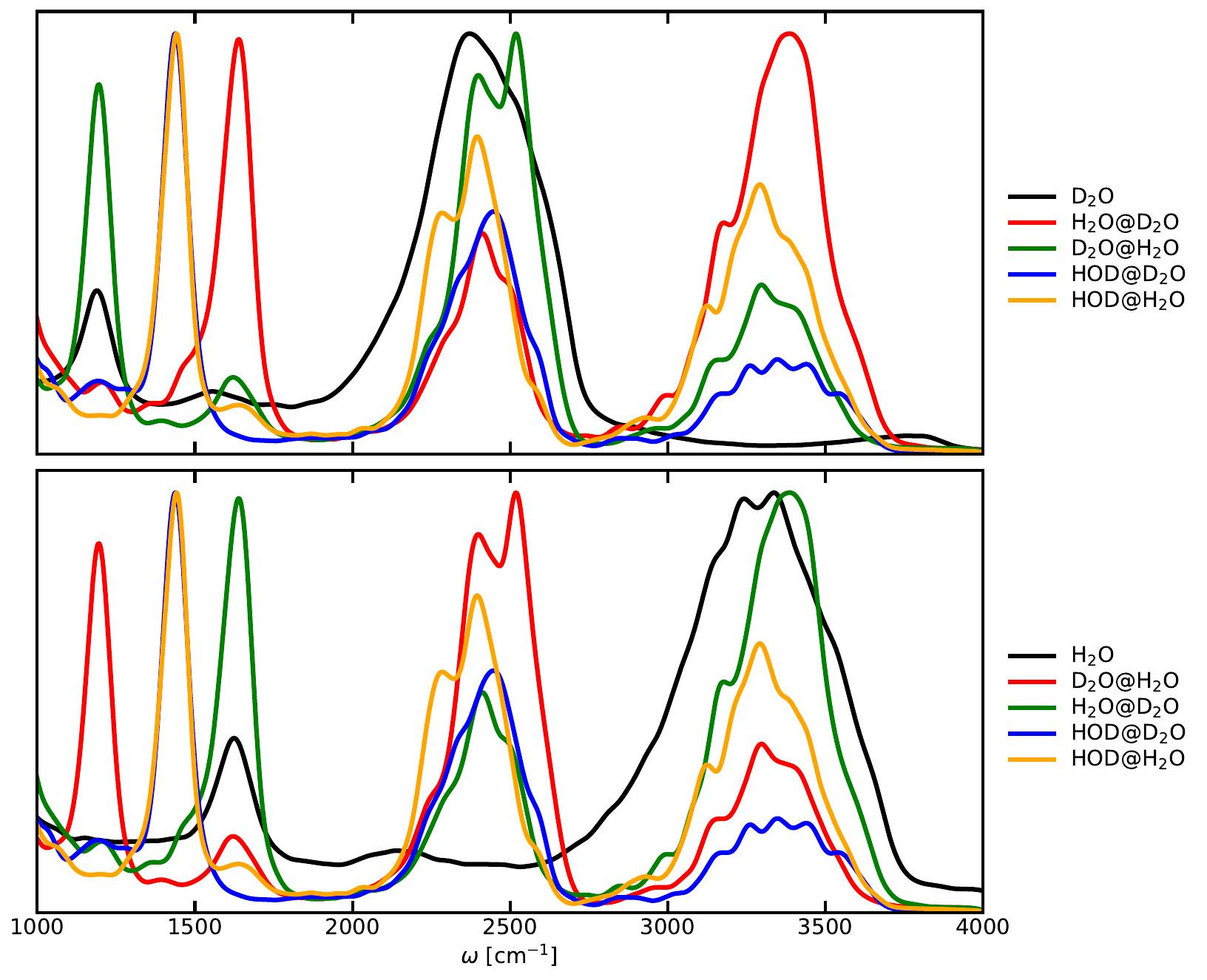}
    \end{tabular}
  \end{center}
  \caption{(Colour online) The infrared spectrum in the isotopic mixtures H$_2$O@D$_2$O,
    D$_2$O@H$_2$O, HOD@D$_2$O and HOD@H$_2$O.}
  \label{figure:IR_spectrum_mixtures}
\end{figure*}

In figure~\ref{figure:IR_spectrum_mixtures} we further show the the IR spectrum evaluated in the various mixtures H$_2$O@D$_2$O, D$_2$O@H$_2$O, HOD@D$_2$O and HOD@H$_2$O. 
As expected,  there appear peaks in the frequency ranges corresponding to both the O-H and O-D stretching and the H-O-H, H-O-D and D-O-D bending frequencies.

%
%

\subsection{Dependence on the coordination in the hydrogen bond network}\label{section:purewater-coordination}

We start the analysis from the average intramolecular O-H bond lengths, shown
in table~\ref{Table:constrained_intramolecular_dOH} in the different HB
coordinations $C_\mathrm{HB}$ on the constrained, central molecule. The main
effects are as expected: an O-H bond becomes elongated when a HB is donated by
this OH group, and more so when there are more accepted HBs on the O since that
weakens the intramolecular O-H bond strength. We have also evaluated the
average dipole moment on the central molecule $\overline{D}$ via localization
on Wannier functions. Here, we clearly see  how the dipole moment increases upon
having more HBs formed except the cases where the molecule accepts three HBs.

\begin{table}[!t]
  \caption{Average intramolecular bond lengths and dipole moment in the
    constrained simulations.}
  \label{Table:constrained_intramolecular_dOH}
  \vspace{1ex}
\begin{center}
\begin{tabular}{c|cc|c}\hline\hline
$C_\mathrm{HB}$ & $d(\mathrm{OH}_a)$ [\AA] & $d(\mathrm{OH}_b)$ [\AA] &
  $\overline{\mu}_\mathrm{W}$ [D] \strut\\\hline
  \labelZZZ & 0.975$\pm$0.024 & 0.975$\pm$0.024 & 2.06\strut\\
  \labelZZO & 0.975$\pm$0.023 & 0.984$\pm$0.024 & 2.23\strut\\
  \labelZOO & 0.981$\pm$0.026 & 0.981$\pm$0.025 & 2.29\strut\\
\hline
  \labelOZZ & 0.976$\pm$0.025 & 0.976$\pm$0.025 & 2.38\strut\\
  \labelOZO & 0.975$\pm$0.021 & 0.991$\pm$0.028 & 2.65\strut\\
  \labelOOO & 0.986$\pm$0.028 & 0.986$\pm$0.028 & 2.78\strut\\
\hline
  \labelTZZ & 0.977$\pm$0.024 & 0.977$\pm$0.024 & 2.46\strut\\
  \labelTZO & 0.976$\pm$0.024 & 0.997$\pm$0.028 & 2.80\strut\\
  \labelTOO & 0.992$\pm$0.028 & 0.992$\pm$0.028 & 3.07\strut\\
\hline
  \labelTOOs& 0.997$\pm$0.029 & 0.997$\pm$0.029 & 3.20\strut\\
  \labelHOO & 0.995$\pm$0.030 & 0.993$\pm$0.027 & 2.97\strut\\
  \labelHnm & 0.991$\pm$0.029 & 0.991$\pm$0.027 & 2.97\strut\\\hline\hline
\end{tabular}
\end{center}
\end{table}

\begin{figure*}[!t]
  \begin{center}
    \begin{tabular}{c}
      \includegraphics[width=0.7\textwidth]{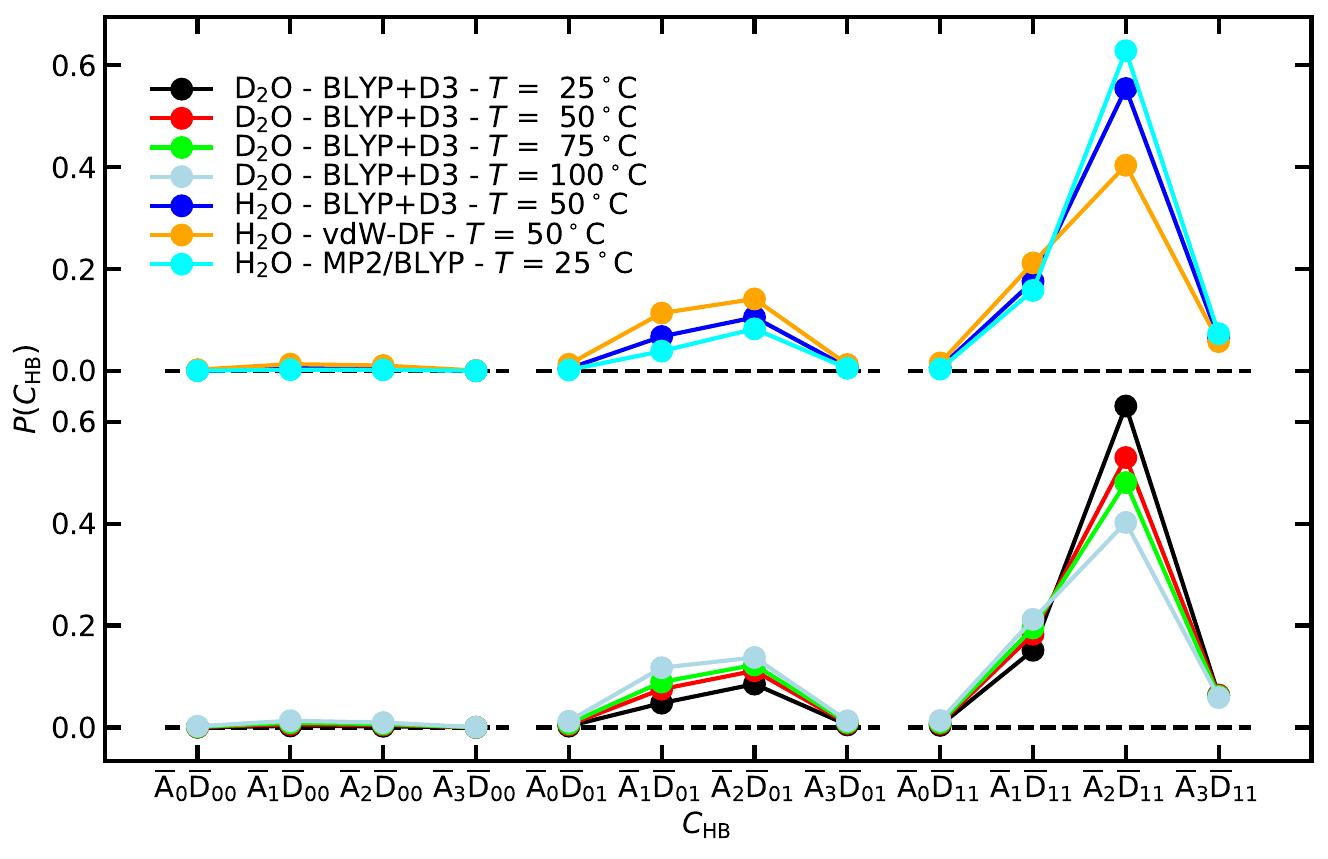}
    \end{tabular}
  \end{center}
  \caption{(Colour online) Distribution of HB configurations $C_\mathrm{HB}$ in the
    unconstrained trajectories.}
  \label{figure:HBcfg}
\end{figure*}

We determine how well the constraint on the HB coordination works by evaluating
the distribution of $C_\mathrm{HB}$; the results are shown in
figure~\ref{figure:HBcfg}b. Overall the central molecule is coordinated as
given by the constraint,\footnote{We remark that the type of constraint applied is not exactly related to the
	criterion of a {HB}, and we remind that we used relatively loose radii in the
	constraint, so as not to affect the vibrations in a given configuration too
	much. {T}herefore, the good agreement between is all the more rewarding.}
yet some
differences occur, in particular in cases where the forcing of a given
coordination leads to a strong difference in the number of acceptor and donor
HBs from the ``standard'' coordination A$_2$D$_{11}$, such as A$_2$D$_{00}$.
The distributions in the unconstrained trajectories, shown in
figure~\ref{figure:HBcfg}a, indicate which coordinations naturally occur in
the water: They are always close to the full coordination A$_2$D$_{11}$,
sometimes having one accepting HB more or less (A$_3$D$_{11}$, A$_1$D$_{11}$)
or one donor HB less (A$_2$D$_{01}$, A$_1$D$_{01}$); configuations with either
no acceptor or donor HBs are very rare.

\begin{figure*}[!t]
  \begin{center}
    \begin{tabular}{c}
      \includegraphics[width=0.55\textwidth]{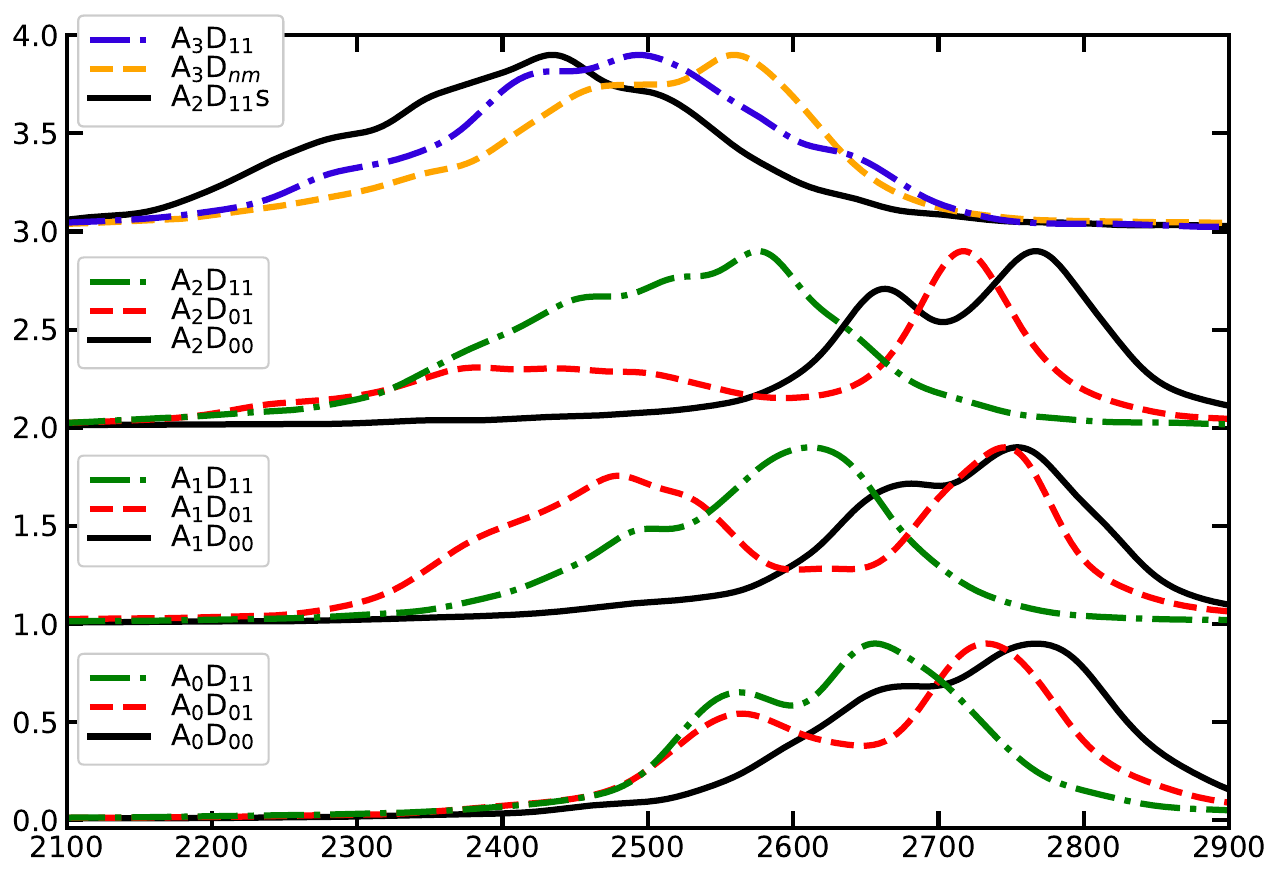}\\%
      \includegraphics[width=0.55\textwidth]{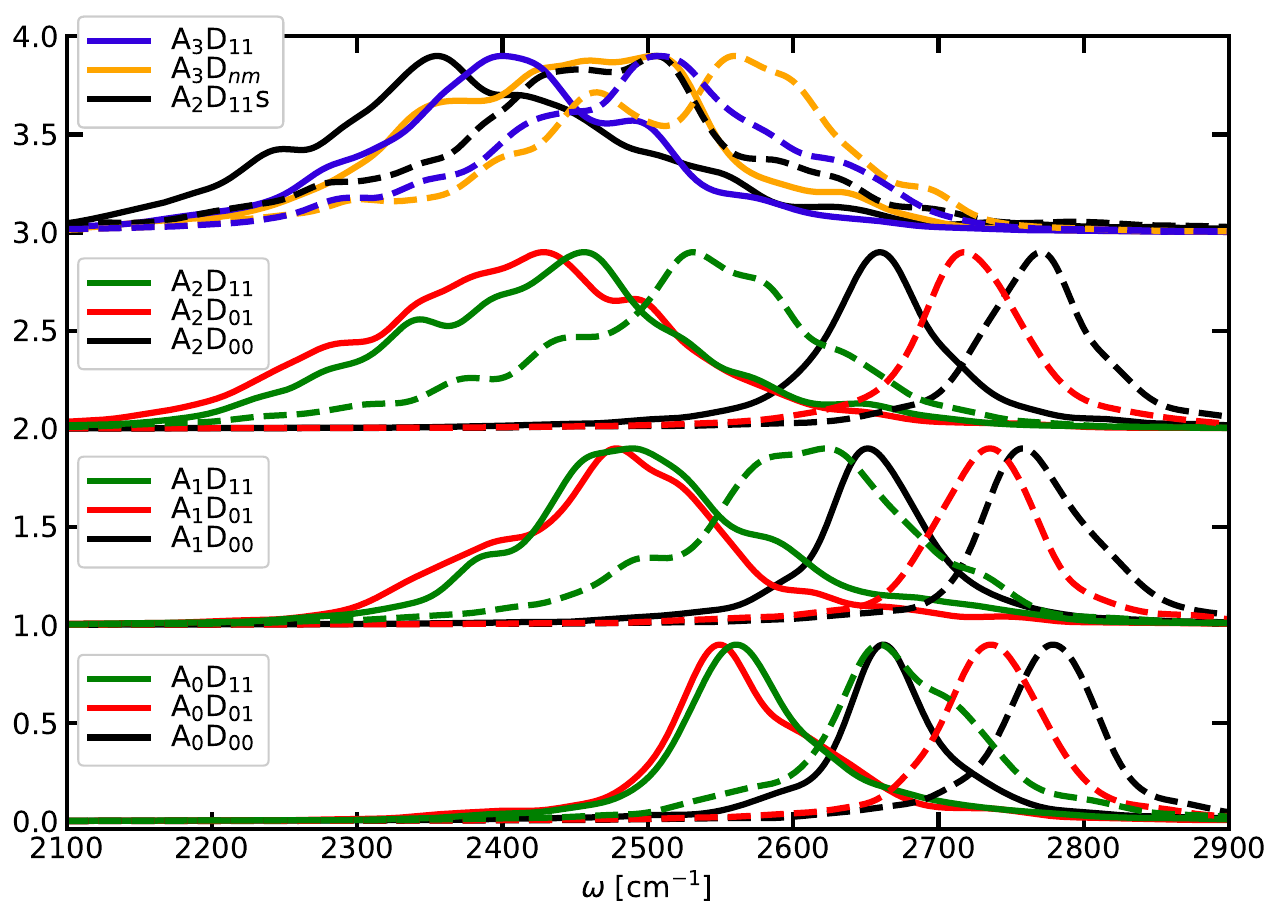}%
    \end{tabular}
  \end{center}
  \caption{(Colour online) The power spectrum on the oxygen atoms  (top panel)  and the
    localized power spectra from ENMA (bottom panel) in the simulations
    with the restrained coordination of the hydrogen bonds.}
  \label{figure:power_spectra_D2O_constrained}
\end{figure*}

The power spectrum of the constrained molecule, displayed in
figure~\ref{figure:power_spectra_D2O_constrained}, gives indication of the
frequency ranges in given coordinations. However, we concentrate on the frequency
distributions yielded via the ENMA; these are also shown in
figure~\ref{figure:power_spectra_D2O_constrained}, and the average frequencies
of the two OH effective stretching modes in
table~\ref{Table:constrained_frequency}.

\begin{table}[!t]
  \caption{Frequencies obtained from localized ENMA vibrations at different
    coordinations $C_\mathrm{HB}$ on the central water molecule.}
  \label{Table:constrained_frequency}
  \vspace{2ex}
\begin{center}
\begin{tabular}{l|ccc|ccc}\hline\hline
$C_\mathrm{HB}$ & \multicolumn{3}{c|}{$\omega$} & \multicolumn{3}{c}{$\Delta\omega$}
  \strut\\\hline
  \labelZZZ & 1234 & 2686 & 2797 & ref & ref & ref\strut\\
  \labelZZO & 1224 & 2587 & 2756 & $-10$ & $-99$ & $-41$\strut\\
  \labelZOO & 1240 & 2628 & 2722 & 6 & $-58$ & $-74$\strut\\\hline
  \labelOZZ & 1242 & 2678 & 2793 & 9 & $-8$ & $-4$\strut\\
  \labelOZO & 1233 & 2500 & 2750 & $-1$ & $-186$ & $-47$\strut\\
  \labelOOO & 1230 & 2531 & 2628 & $-4$ & $-155$ & $-168$\strut\\\hline
  \labelTZZ & 1260 & 2678 & 2788 & 26 & $-8$ & $-9$\strut\\
  \labelTZO & 1246 & 2428 & 2744 & 12 & $-258$ & $-52$\strut\\
  \labelTOO & 1226 & 2449 & 2529 & $-8$ & $-237$ & $-268$\strut\\\hline
 \labelTOOs & 1224 & 2431 & 2498 & $-10$ & $-255$ & $-298$\strut\\
  \labelHOO & 1204 & 2453 & 2529 & $-30$ & $-233$ & $-268$\strut\\ 
  \labelHnm & 1204 & 2453 & 2529 & $-30$ & $-233$ & $-268$\strut\\\hline\hline
\end{tabular}
\end{center}
\end{table}

A similar analysis is given in figure~S16, where we have calculated the
Fourier transform of the fluctuations of the intramolecular $d_\mathrm{OH}$
bond length and their average and difference, the latter mimicking a symmetric
and asymmetric effective mode. These frequency distributions reflect well the
ENMA modes, with a clear dual-peak structure in individual vibrations when 
one of the donor HBs is present and the other one is missing, yielding a breaking of
the symmetry from the symmetric and asymmetric effective modes; correspondingly,
the two peaks are better discerned in the average and difference of
$d_\mathrm{OH}$ when both donor HBs are either present or broken; there, the
symmetric and asymmetric effective modes provide a consistent picture of the
local vibrations. 

\begin{figure}[!t]
\begin{center}
\begin{tabular}{c}
\includegraphics[width=0.6\columnwidth]{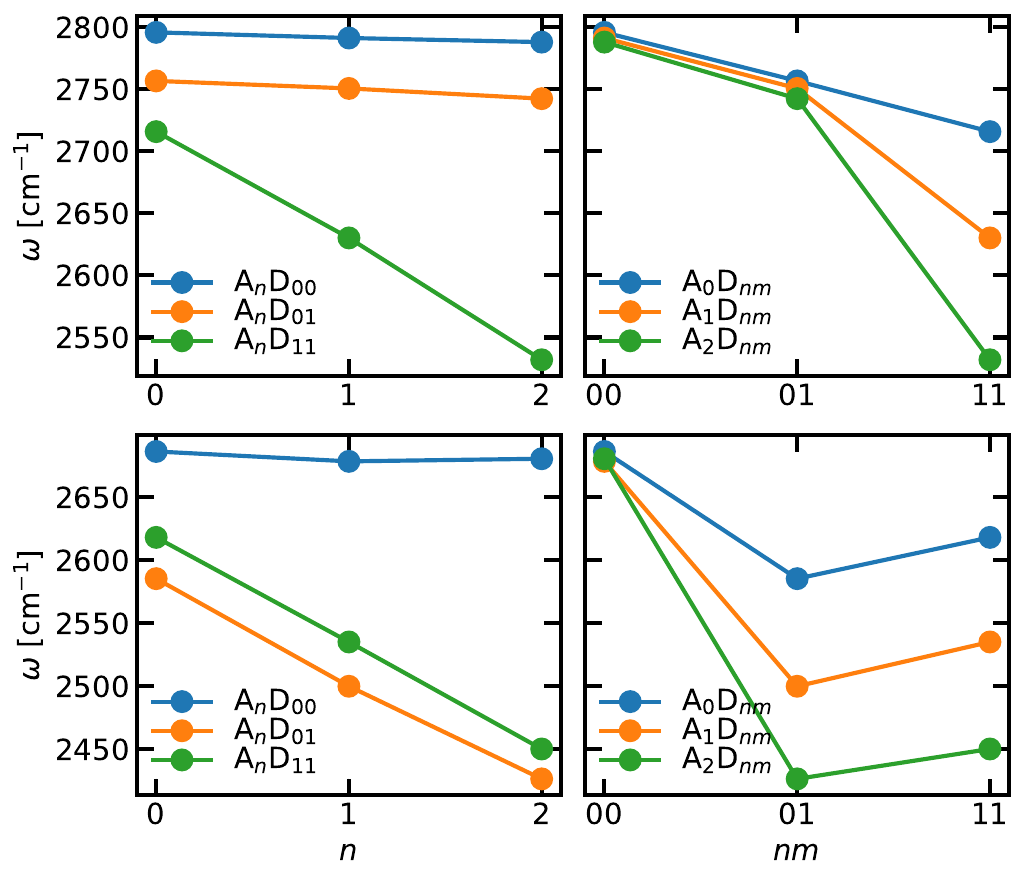}
\end{tabular}
\end{center}
\caption{(Colour online) Dependence of the two ENMA O-D stretching frequencies of a water molecule in D$_2$O in restricted coordinations upon a varying
  number of local number of acceptor and donor HBs.}
\label{figure:coordination_enma_frequencies_D2O}
\end{figure}

Against these observations it is easier to comprehend the changes upon varying
coordination shown in the
figure~\ref{figure:coordination_enma_frequencies_D2O}: When going from a quasi-free molecule $C_\mathrm{HB} = $A$_n$D$_{00}$ to A$_n$D$_{01}$ and to A$_n$D$_{11}$,
the lower of the ENMA frequencies has its minimum in the middle configuration.
This is probably caused by the collapse of the symmetric-asymmetric effective
modes into two independent vibrations rather than an intrinsic lowest bond
strength in those configurations.

We still make an observation of the hydrogen bond using the O-O-RDF split into
contributions from hydrogen bonded and non-hydrogen bonded molecular pairs;
the distributions are shown in figure~S2. 
They appear as expected, with the major part of the first peak in the RDFs
arising from hydrogen bonded molecules, and the non-hydrogen bonded pairs
appearing at about 3.0~\AA{} in the constrained samples and at about 2.7~\AA{}
in the normal trajectories. The amount of non-hydrogen bonds increases in
D$_2$O upon increasing temperature, indicating more molecules penetrating into
the first hydration shell of the molecules; we conclude that these are not
hydrogen bonded to the central molecule since the share of the molecules with
three acceptor HBs does not change in figure~\ref{figure:HBcfg}a.


%
%
\section{Discussion}\label{section:discussion}

First of all, we remark that our overall description of the IR and Raman
spectra of liquid water is relatively good: the main features and trends known
from experiemets are very well reproduced qualitatively.

From earlier studies it has become apparent that the BLYP+D3 approach yields
relatively realistic description of liquid water. Yet short-comings persist,
like the too high equilibrium density at ambient
conditions~\cite{DelBen_2015_a}, the modest over-structuring of the
liquid~\cite{Jonchiere_2011_a} arising due to a too high  melting
temperature~\cite{Seitsonen_2016_a} evidenced also in the collective
vibrations~\cite{Bryk_2016_a}. 

When we compare our work to the previous literature on Raman spectra from structure
method-based MD simulations~\cite{Wan_2013_a}, the agreement is reasonable, but
we note that highly elevated temperature was used~\cite{Wan_2013_a} in
order to reduce the known tendency of
over-structuring~\cite{Kuo_2004_JPCB_a,Lin_2012_a} of the PBE treatment of XC.



Our simulated IR spectrum is very close to the one obtained earlier by
Silvestrelli and Parrinello~\cite{Silvestrelli_1997_CPL_a}. 
The reduction in the dipole moment upon an increased temperature is in
line with the simulations of Guillot and Guissani~\cite{Guillot_1995_a}. The
shifts in the shape of the Raman spectra are in line with those from the
experiments and the neural networks~\cite{Sommers_2020_PCCP_a}.


In table~\ref{fig:scaled_frequency} we compare the shifts in the OH/OD
stretching frequency, defined as $\omega/\omega_\mathrm{ref}-1$, from our
simulations in the D$_2$O with the constrained coordination to those obtained
in \cite{Auer_2007_PNAS_a} in HOD in D$_2$O. We note that the methods used
are somewhat different, because in our simulations we obtain the frequencies from
the analysis with the method ENMA, whereas in~\cite{Auer_2007_PNAS_a} the
Authors use a semi-classical method, with the quantized dynamics on the single
O-H bond; yet our goal is to quantify the trends on the shifts in the
vibrational frequencies. Also in~\cite{Auer_2007_PNAS_a}, the instantaneous
configurations are grouped differently, where both the configurations labelled
with $N_\mathrm{O}=1$ correpond to our cases A$_\mathrm{0}$D$_\mathrm{nm}$ and
A$_\mathrm{0}$D$_\mathrm{nm}$. As the $\omega_\mathrm{ref}$, we use the value
in the coordination A$_0$D$_{00}$. We see the correspondence in the shifts, in
particular, between the configurations A$_1$D$_{11}$ and 3$_\mathrm{D}$ and
A$_2$D$_{11}$ and 4$_\mathrm{D}$, and somewhat less clearly between the cases
A$_1$D$_{01}$ and 2$_\mathrm{SH}$. However, in the asymmetric coordinations
A$_n$D$_{01}$, the two simulations  differ more than in~\cite{Auer_2007_PNAS_a} where the central molecule is already asymmetric, namely
HOD.

\begin{table}[!t]
  \caption{Relative difference in the OH/OD stretching frequency at different
    coordinations of the hydrogen bonds both from our simulations and those from~\cite{Auer_2007_PNAS_a}.}
  \label{fig:scaled_frequency}
  \vspace{2ex}
\begin{center}
\begin{tabular}{c|cc|cc|cccc|c}
\hline\hline
Coordination & \multicolumn{2}{c|}{$\omega$} & \multicolumn{2}{c|}{shift} & $N_\mathrm{O}$ & $N_\mathrm{H}$ & $N_\mathrm{D}$ & label & shift\strut\\\hline
 A$_0$D$_{0  0}$ & 2672 & 2786 &  ref & ref & \strut\\
 A$_1$D$_{0  0}$ & 2700 & 2805 &  0.010 & 0.007 & \raisebox{0.9ex}[0pt][0pt]{1} & \raisebox{0.9ex}[0pt][0pt]{0} & \raisebox{0.9ex}[0pt][0pt]{0} & \raisebox{0.9ex}[0pt][0pt]{1$_\mathrm{N}$} & \raisebox{0.9ex}[0pt][0pt]{$-$0.014}\strut\\
 A$_2$D$_{0  0}$ & 2674 & 2778 &  0.001 & $-$0.003 & 2 & 0 & 0 & 2$_\mathrm{N}$ & $-$0.026\strut\\\hline
 A$_0$D$_{0  1}$ & 2587 & 2743 & $-$0.032 & $-$0.015 \strut\\
 A$_1$D$_{0  1}$ & 2512 & 2750 & $-$0.060 & $-$0.013 & 1 & 0 & 1 & 2$_\mathrm{SD}$ & $-$0.011\strut\\
                 &      &     &          &     & 1 & 1 & 0 & 2$_\mathrm{SH}$ & $-$0.067\strut\\
 A$_2$D$_{0  1}$ & 2588 & 2680 & $-$0.031 & $-$0.038 & 2 & 0 & 1 & 3$_\mathrm{SD}$ & $-$0.022\strut\\
                 &      &     &          &     & 2 & 1 & 0 & 3$_\mathrm{SH}$ & $-$0.088\strut\\\hline
 A$_0$D$_{1  1}$ & 2579 & 2686 & $-$0.035 & $-$0.036 & \strut\\
 A$_1$D$_{1  1}$ & 2497 & 2590 & $-$0.065 & $-$0.071 & \raisebox{0.9ex}[0pt][0pt]{1} & \raisebox{0.9ex}[0pt][0pt]{1} & \raisebox{0.9ex}[0pt][0pt]{1} & \raisebox{0.9ex}[0pt][0pt]{3$_\mathrm{D}$} & \raisebox{0.9ex}[0pt][0pt]{$-$0.063}\strut\\
 A$_2$D$_{1  1}$ & 2457 & 2539 & $-$0.080 & $-$0.089 & 2 & 1 & 1 & 4$_\mathrm{D}$ & $-$0.083\strut\\
\hline\hline
\end{tabular}
\end{center}
\end{table}

We note and admit here the admissions and unnecessary approximations and
shortcuts in our simulations: in addition to the issues related to the
approximative exchange-correlation functionals, neglected nuclear quantum
effects, classical dynamics instead of the quantized dynamics, we also used a
constant, empirical density of water instead of letting it adjust to the value
that we would reach if the isothermal-isobaric ensemble were employed;
indeed a density of 1.066~g/cm$^3$~\cite{DelBen_2013_JPCL_a} has beeen found
under the ambient conditions.

%
%
\section{Conclusions}\label{section:conclusions}

In this study we have performed DFT-based MD simulations of pure liquid
water. We  collected data on various structural and dynamical quantities,
and we hope that the results will be useful in the interpretation of future
investigations.

In particular, we investigated what happens to the O-H bond lengths, dipole
moments and the vibrational frequencies when the coordination of the hydrogen
bonds around the central molecules is different.  We  achieved this by
explicitly restraining the distance of the neighbouring atoms from the oxygen
and the two hydrogen or deuterium nuclei of the central molecule. We find
definitive shifts in these quantities, which form  the basis for the
observations made in the normal, unconstrained simulations and thus further in
the experimental results.

In the simulations of the liquid without constraints, we obtain an insight into the
various underlying HB configurations in the course of simulations by
correlating the characteristic frequency domain of the different HB
coordination to the full spectrum in the fluctuating liquid. We can  see
how the numbers of the HBs reduce with the increased simulated temperature, yet
interestingly the number of the configurations with five HBs --- with our
specific definition of an HB at least --- is reduced only slightly whereas the
average number of the HBs drops clearly.

In summary, we have managed to contribute somewhat further to the disclosure of the
``secrets'' of the hydrogen bond network and how the effects are detected in
the experimental results.

%
%
\section*{Acknowledgements}\label{section:acknowledgements}

We acknoledge Mauro Del Ben, Joost VandeVondele and J\"{u}rg Hutter for
providing us with the MP2 trajectories, and Marcella Iannuzzi for the numerous fruitful
discussions and support over the years.

\newpage
\ukrainianpart

\title{Спектроскопія коливних станів води: температурні та координаційні ефекти в комбінаційних та інфрачервоних спектрах}
\author{Р. Вулем'є\refaddr{ENS}, A. П. Сейтсонен\refaddr{UZH,ENS}} 

\addresses{
	\addr{ENS} Науковий підрозділ з вивчення процесів вибіркової активації на основі одноелектронної або радіаційної передачі енергії (PASTEUR), відділення хімії, Нормальна вища школа, Паризький дослідницький університет наук і літератури (PSL), Сорбоннський університет, Національний центр наукових досліджень (CNRS), F-75005 Париж, Франція
	\addr{UZH} Інститут хімії, Цюріхський університет,
	Вінтертурештрассе 190, CH-8057 Цюріх, Швейцарія
}

\makeukrtitle

\begin{abstract}
	Вода є найпоширенішою рідиною, що володіє багатьма екзотичними та аномальними властивостями. 
	Незважаючи на очевидну просту хімічну формулу, її здатність утворювати динамічну сітку 
	водневих зв'язків призводить до виникнення різноманітних фізичних ефектів. 
	Досліджуються коливання у воді на основі молекулярної динаміки
	з акцентом переважно на сигнатурах комбінаційних та інфрачервоних спектрів. 
	Вивчено вплив температури на коливні спектри; введено деталі координаційних співвідношень між вод\-не\-ви\-ми зв’язками на основі моделювання з енергетичними обмеженнями для отримання кількісної інформації про залежність частот сусідніх молекул. 
	Також розглянуто відмінності, що виникають при використанні різних методів розрахунку електронної структури для обчислення 
	сил, що діють на іони. Наведено результати для кутових кореляцій, для ізотопних сумішей HOD в H$_2$O/D$_2$O та діелектричних сталих води.
	\keywords вода, комп'ютерне моделювання, коливна спектроскопія, молекулярна динаміка, метод функціоналу густини, 
	водневі зв'язки
\end{abstract}




  \end{document}